%% file: ms.tex
\pdfoutput=1 
\documentclass{llncs}
\usepackage{graphicx}
\usepackage{subcaption}
\usepackage[utf8]{inputenc}
\usepackage{xspace}
\usepackage{makecell}
\usepackage{xcolor}
\usepackage{listings}
\usepackage{amsmath}
\usepackage{comment}
\usepackage{float}
\usepackage[colorlinks=true,breaklinks=true]{hyperref}
\usepackage[noabbrev]{cleveref}
\input{macros}

\makeatletter
\def\orcidID#1{\smash{\href{http://orcid.org/string#1}{\protect\raisebox{-1.25pt}{\protect\includegraphics{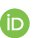}}}}}
\makeatother
\begin{document}

\title{Protocol-based Smart Contract Generation}
%
%
\author{Afonso Falcão\inst{1} \and
Andreia Mordido\inst{1}\orcidID{0000-0002-1547-0692} \and
Vasco T. Vasconcelos\inst{1}\orcidID{0000-0002-9539-8861}}
\authorrunning{A. Falcão, A.Mordido, and V. Vasconcelos}
%
\institute{LASIGE, Faculty of Sciences, University of Lisbon, Portugal\\
\email{anfalcao@lasige.di.fc.ul.pt}
\email{afmordido@ciencias.ulisboa.pt}
\email{vmvasconcelos@ciencias.ulisboa.pt}
}%
\maketitle              
\begin{abstract}
  The popularity of smart contracts is on the rise, yet breaches in reliability
  and security linger. Among the many facets of smart contract reliability, we
  concentrate on faults rooted in out-of-order interactions with contract
  endpoints. We propose \ourlanguage, a protocol language to describe valid
  patterns of interaction between users and endpoints. \ourlanguage not only
  ensures correct interactive behaviour but also simplifies smart contract
  coding. From a protocol description, our compiler generates a smart contract
  that can then be completed by the programmer with the relevant business logic.
  The generated contracts rely on finite state machines to control endpoint
  invocations. As a proof of concept, we target Plutus, the contract programming
  language for the Cardano blockchain. Preliminary evaluation points to a 75\%
  decrease in the size of the code that developers must write, coupled with an
  increase of reliability by enforcing the specified patterns of interaction.

\end{abstract}
\input{intro}
\input{motivation}

\input{scribble_contract}
\input{contract_gen}

\input{evaluation}
\input{conclusion_futurework}
\bibliographystyle{splncs04}
\bibliography{ms}
\appendix

\input{guessing-game-code}
\input{logs}
\input{scribble-protocols}
\input{plutus-code}
\end{document}

%% file: macros.tex
\definecolor{darkgreen}{rgb}{0,0.4,0.2}
\definecolor{darkblue}{rgb}{0.1,0.1,0.9}
\definecolor{darkgrey}{rgb}{0.5,0.5,0.5}
\definecolor{orangered}{RGB}{239,134,64}
\definecolor{codepurple}{rgb}{0.65,0.2,0.8}
\lstdefinestyle{eclipse}{
  breaklines=true,
  basicstyle=\sffamily\footnotesize,
  emphstyle=\color{red}\bfseries,
  keywordstyle=\color{codepurple}\bfseries,
  commentstyle=\color{darkgreen},
  stringstyle=\color{orangered},
  numberstyle=\color{darkgrey},
  emphstyle=\color{red},
  keywordstyle = [2]{\color{darkblue}},
  morecomment=[s][\color{darkgreen}]{/**}{*/},
  morecomment=[l]{//},
  showstringspaces=false,
  tabsize=2
}

\lstdefinelanguage{scribble}{
  language=Haskell,
  style=eclipse,
  morekeywords={role,protocol,choice,rec,trigger,from, save, slot, funds, do ,
    interrupt, field, val,ByteString, HashedString, Value},
  showstringspaces=false,
  breaklines=true,
  tabsize=2
}

\newcommand{\ourlanguage}[0]{\textsc{SmartScribble}\xspace}


%% file: intro.tex
\section{Introduction}
\label{sec:intro}

Smart contracts are a focal point of modern blockchain environments. Such
contracts were firstly popularized by Ethereum~\cite{Buterin2014}, but soon 
thereafter other networks developed their own smart contract languages, 
enabling the implementation of blockchain-based decentralized applications
between untrusted parties.

Smart contracts usually operate over user owned assets, thus,
vulnerabilities in programs and in the underlying programming languages can lead
to considerable losses. The famous attack on the DAO resulted in a theft of
approximately 60 million USD worth of
Ether~\cite{Atzei2017,DAOExplained,Chen2020}. Due to recent exploitations of
vulnerabilities in smart contracts, blockchain providers
turned their attention to the development of robust programming languages, often
relying on formal verification, including Liquidity\footnote{\url{http://www.liquidity-lang.org/doc/index.html}} by Tezos,
Plutus by IOHK~\cite{Brunjes}, Move by Facebook~\cite{blackshear2019move}, and
Rholang\footnote{\url{https://github.com/rchain/rchain/tree/master/rholang-tutorial}} 
by RChain.
Such languages
aim at offering flexible and complex smart contracts while assuring that
developers may fully trust contract behaviour. Unfortunately, for Plutus, the 
last objective
has not been completely achieved yet. As we show in the next section,
in Plutus, assets can be easily lost forever to the ledger with a simple unintended
interaction.

To counter unplanned interactions with smart contracts endpoints while automating the
development of boilerplate code, we propose \ourlanguage, a protocol
specification language for smart contracts.
Its syntax is adapted from the
Scribble protocol language~\cite{Yoshida2014} to the smart contract trait and
features primitives for sequential composition, choice, recursion, and
interrupts. 
Protocols in \ourlanguage specify interactions between participants
and the ledger, as well as triggers to interrupt protocol execution. The
business logic underlying the contract can be added by the programmer after the
automatic generation of the smart contract boilerplate code. The generated
code relies on a finite state machine to validate all interactions,
precluding unexpected behaviours.

\ourlanguage currently targets Plutus, a native smart contract programming
language for the Cardano blockchain~\cite{Kiayias2017}, based on the Extended
Unspent Transaction Output model~\cite{DBLP:conf/fc/Chakravarty0MMJ20}, a solution 
that expands the limited expressiveness of the Unspent Transaction Output model. 
In UTxO, transactions consist of a list of inputs and outputs. Outputs correspond to 
the quantity available to be spent by inputs of subsequent transactions.
Extended UTxO expands UTxO's expressiveness without switching to 
an account-based model, that introduces a notion of shared mutable state,
ravelling contract semantics~\cite{DBLP:conf/fc/Chakravarty0MMJ20}.
Nevertheless, the framework we propose can be integrated with other smart
contract languages and blockchain infrastructures expressive enough to support state machines.

Several works have been adopting state machines to control the interaction of 
participants with smart contracts.
FSolidM~\cite{Mavridou2017}---the closest proposal to \ourlanguage---introduces a
model for smart contracts based on finite state machines. FSolidM relies on the
explicit construction of finite state machines for contract specification;
instead, we automatically generate all state machine code.
On a different fashion, the model checker Cubicle~\cite{DBLP:conf/fm/ConchonKZ19} 
encodes smart contracts and the transactional model of the blockchain as a state machine.

\ourlanguage distinguishes itself from other domain-specific languages---BitML,
integrated with the Bitcoin blockchain~\cite{Bartoletti2018},
Obsidian~\cite{coblenz2020obsidian}, a typestate-oriented language, and
Nomos~\cite{nomos}, a functional (session-typed) language---by abstracting the 
interactive behaviour and details of the target programming language through a protocol specification,
only relying on the smart contract language to implement the business logic and thus flattening
the learning curve.

The next section motivates \ourlanguage via an example where assets are lost to
the ledger; \cref{sec:protocols} presents the protocol language and
\cref{sec:generation} contract generation from protocols. \Cref{sec:evaluation}
describes some preliminary results of our evaluation of \ourlanguage, and \cref{sec:conclusion} concludes the paper
and points to future work. \Cref{sec:guessgame_code} contains the source code for the vulnerable contract we explore in our motivation, \cref{sec:logs} presents input and logs for playground simulations, \cref{sec:ap_protocols} contains the definition of \ourlanguage protocols used in \cref{sec:evaluation}, and \cref{sec:plutus_code} the source code for the business logic of our running example.


%% file: motivation.tex
\section{Smart contracts can go wrong}
\label{sec:motivation}

This section identifies a weakness of the Plutus smart contract programming
language. 
Although Plutus is developed with a clear focus on reliability, it lacks
mechanisms to enforce correct patterns of interactions.
\begin{figure}[t!]
  \centering
  \includegraphics*[width=.85\linewidth]{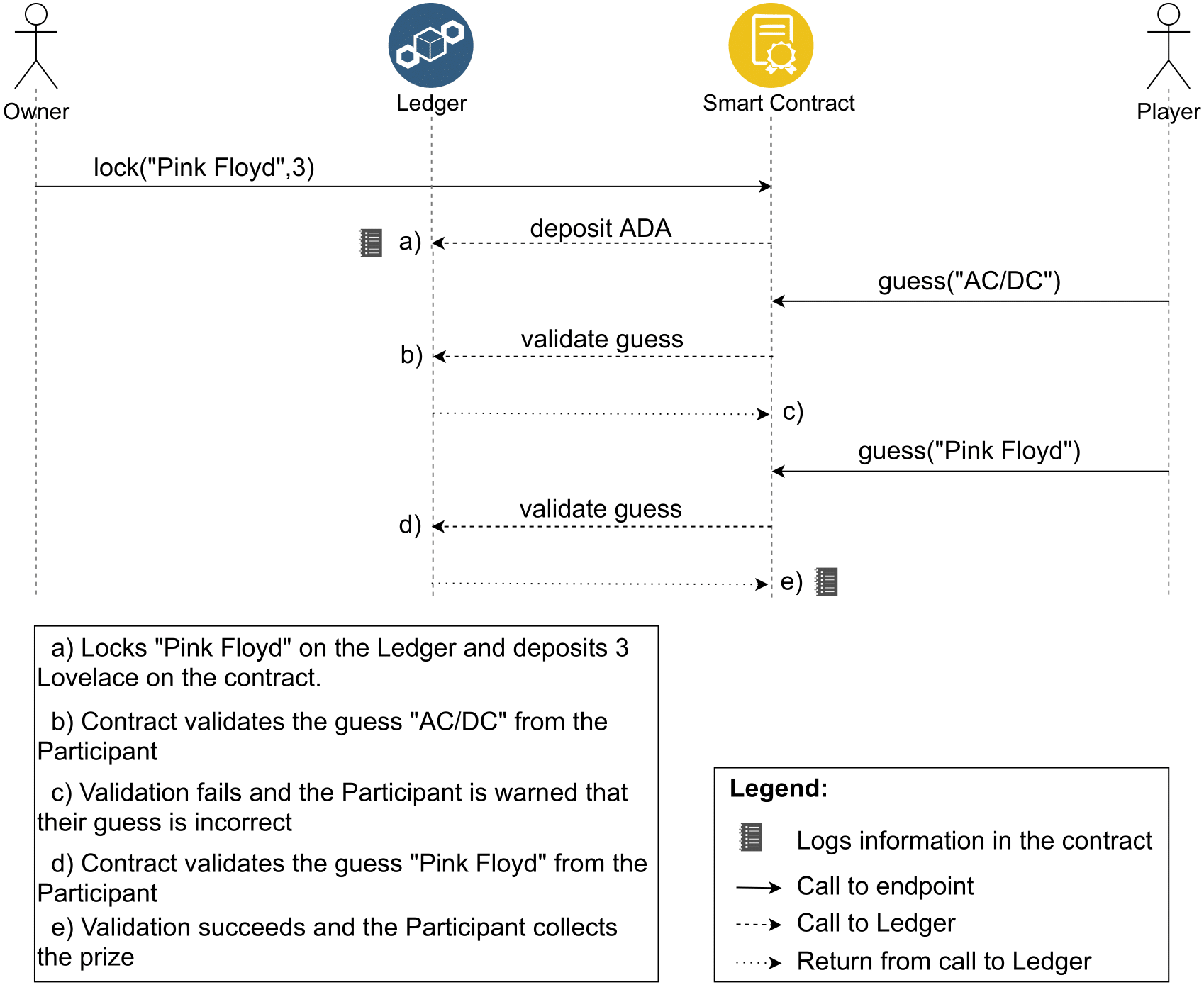}
  \caption{Diagram of a particular well-behaved flow of operations}
  \label{One_Lock_GuessGame}
  \vspace*{-2mm}
\end{figure}
As a running example we consider the popular guessing
game (contract in~\cref{sec:guessgame_code}), the paradigm for secret-based contracts
where participants try to guess a secret and get rewarded if successful. Another
example that falls in this category is a lottery. \Cref{One_Lock_GuessGame}
represents a correct sequence of events in the guessing game:
\begin{enumerate}
	\item The \emph{owner} of the contract locks a secret and deposits a prize in ADA\footnote{ADA is the digital currency of the Cardano blockchain. 1 ADA = 1,000,000 Lovelace.} to be retrieved by the first player who correctly guesses the secret.
	\item The \emph{player} tries to guess the secret.
	\item If the guess matches the secret, the player retrieves the prize and
      the game ends; otherwise, the player is warned that the guess did not
      succeed and the game continues.
\end{enumerate}

\begin{figure}[t!]
  \centering
  \includegraphics*[width=.85\linewidth]{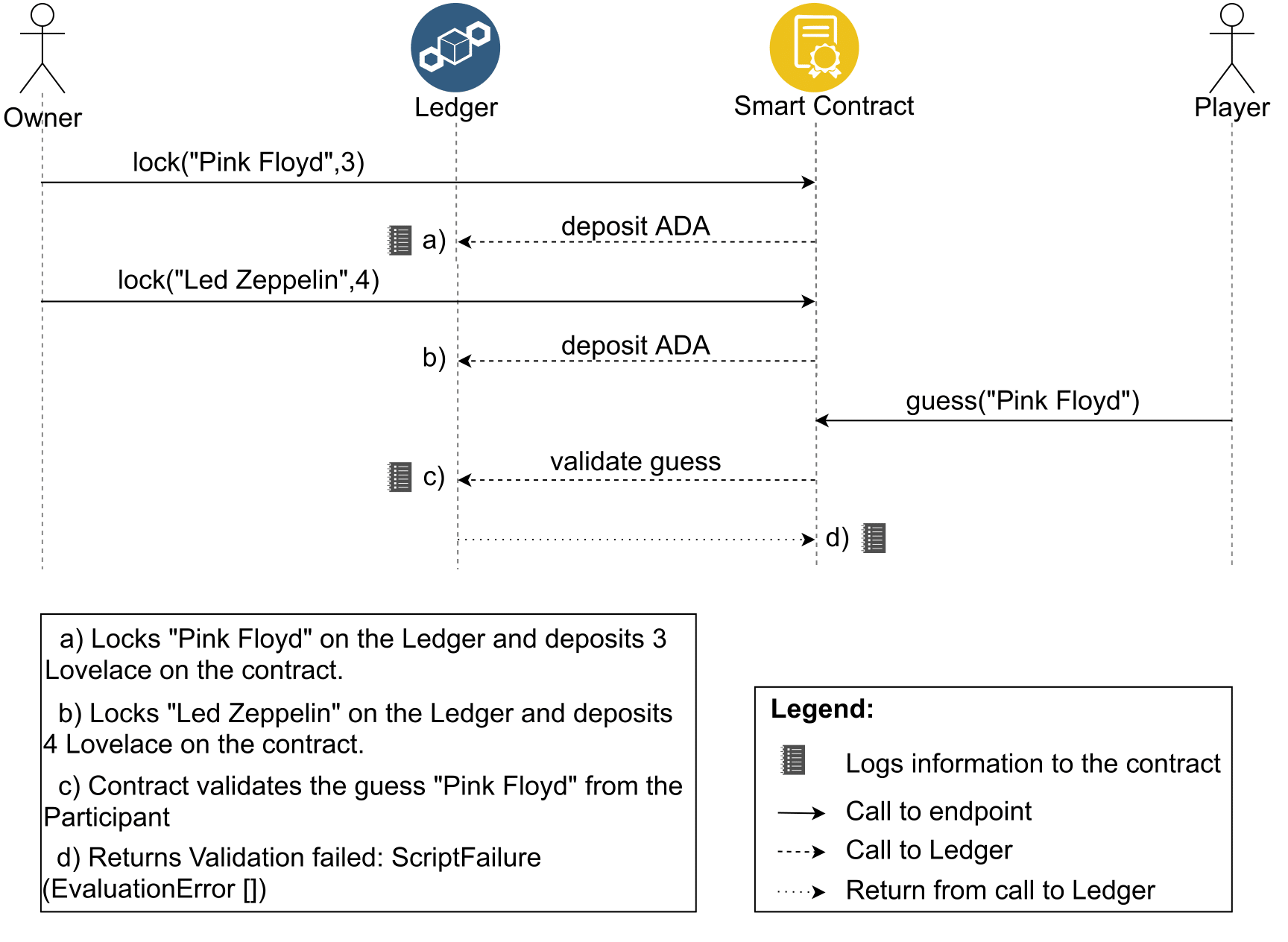}
  \caption{The owner (incorrectly) locks twice and a correct guess results in a failure}
  \label{Two_Lock_GuessGame}
  \vspace*{-2mm}
\end{figure}

\begin{figure}[t!]
	\centering
	\includegraphics*[width=\linewidth]{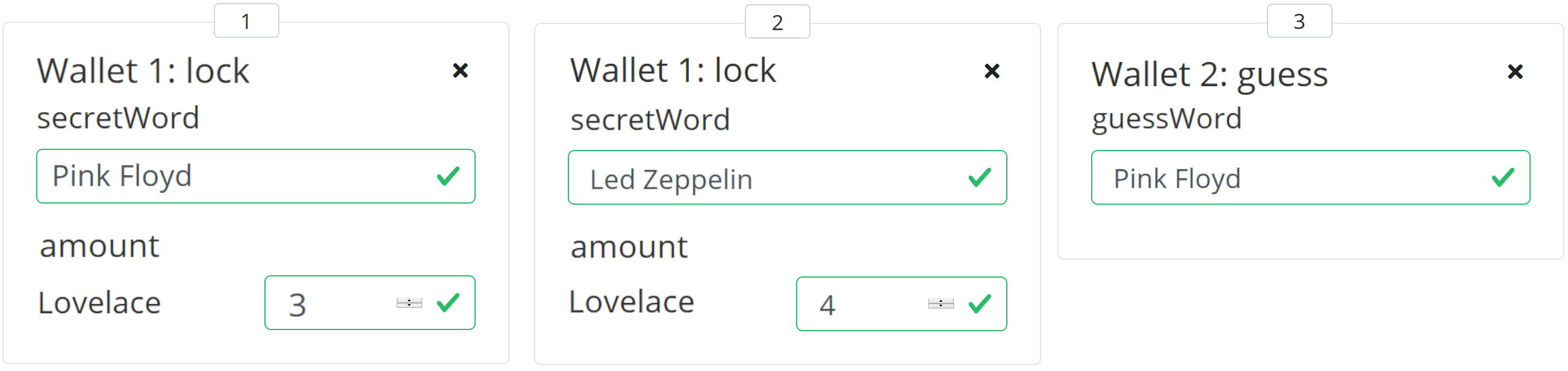}
	\caption{Playground setup for Fig.~\ref{Two_Lock_GuessGame}: the owner makes two consecutive locks, the player guesses the first secret (complete version in Fig.~\ref{Input_TwoLockPlayground_Complete},~\cref{sec:logs})}
	\label{Input_TwoLockPlayground_WithoutSM}
\end{figure}

\begin{figure}[t!]
  \centering
  \includegraphics*[width=.9\linewidth]{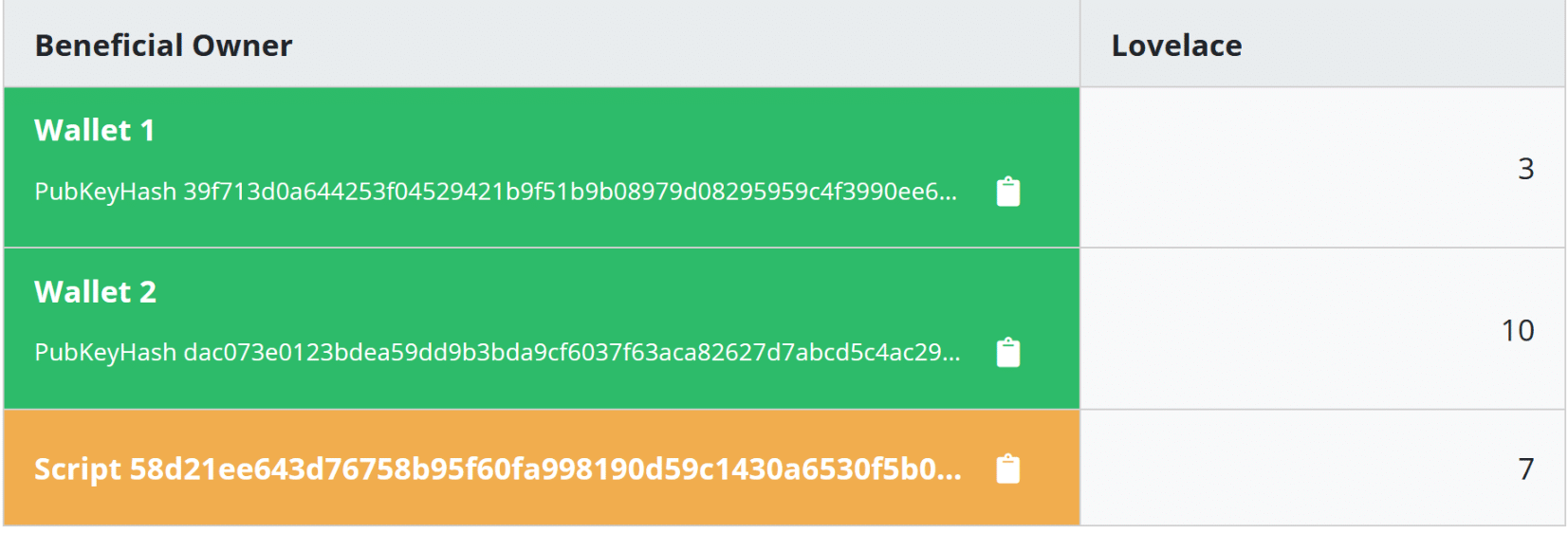}
  \caption{Final balances for input in Fig.
    \ref{Input_TwoLockPlayground_WithoutSM} (Plutus Playground)}
  \label{Output_TwoLockPlayground_WithoutSM_Balances}
  \vspace*{-2mm}
\end{figure}

In Plutus, the parties involved in the protocol are not required to follow valid
patterns of interaction. We explore a scenario where one of the parties deviates
from the (implicitly) expected flow and show that this leads to a faulty
behaviour that is silenced by the blockchain.
\Cref{Two_Lock_GuessGame} represents a scenario where the owner incorrectly
executes two consecutive lock operations and the player provides a correct
guess. A simulation of this scenario in the Plutus
Playground\footnote{\url{https://prod.playground.plutus.iohkdev.io/}} is illustrated
in \Cref{Input_TwoLockPlayground_WithoutSM}.
Starting with 10 Lovelace in both the owner's and player's wallets, we reach
a situation
where the log identifies a validation failure for a guess that coincides with
that of the first lock (see \Cref{Output_TwoLockPlayground_WithoutSM_Logs} in~\Cref{sec:logs}).
The final balances in~\Cref{Output_TwoLockPlayground_WithoutSM_Balances} show 
that the player did not collect the reward despite having
guessed the first secret and the owner lost their money to the contract.

This behaviour is certainly unexpected: rather than overriding the first lock or
having the second lock fail, both secrets are stored in the ledger as outputs.
When a guess is performed, both stored outputs are compared to the guess and, as
a consequence, no guess will ever validate against two different secrets.
Furthermore, the prize is irrevocably lost to the contract without any
possibility of retrieval. This is an unexpected band of silent behaviour that we
want to prevent. Even in this simple scenario 
users lose assets to the ledger. Similar situations are very likely
to occur in complex contracts, with devastating results.


We propose specifying the interaction behaviour of smart contracts through
protocols that describe the valid patterns of interactions between different
classes of users and the contract. Our approach prevents unexpected contract
behaviours by having contracts automatically validating interactions. The
protocol for the guessing game described at the end of next section, detects 
and avoids further attempts to lock secrets, among other unintended interactions.
Note that this type of vulnerability is different from
Transaction Ordering Dependence~\cite{DBLP:journals/access/SayeedMC20} 
that is related to corrupt miners maliciously changing the order of transactions, 
and not the order in which the endpoints are called.

%% file: scribble_contract.tex
\section{Specifying smart contract protocols in \ourlanguage }
\label{sec:protocols}

Scribble~\cite{Yoshida2014} is a language to describe application-level
protocols for communicating systems. It comes with tools  to generate Java or
Python APIs on which developers can base correct-by-construction implementations
of protocols.

\ourlanguage is based as much as possible on Scribble, even if it covers only a
fragment of the language and includes support for smart contract specific
features. The base types of \ourlanguage include
\lstinline[morekeywords={String}]|String|,
\lstinline[morekeywords={HashedString}]|HashedString| (strings stored in the
ledger), \lstinline[morekeywords={PubKeyHash}]|PubKeyHash| (wallet public key
identifiers) and \lstinline[morekeywords={Value}]|Value| (an amount in ADA).
To watch \ourlanguage in action we start with a very simple version of the guessing
game protocol and gradually make it more robust.
Our first version is the straight line guessing game, featuring a sequential composition of three endpoints: \lstinline|lock|, \lstinline|guess|, \lstinline|closeGame|.
\begin{lstlisting}
protocol StraightLineGuessingGame (role Owner, role Player) {
  field HashedString; // save the secret in the contract
  // the owner locks the secret and deposits a prize
  lock (String, Value) from Owner;
  guess (String) from Player;  // the player makes a guess
  // the owner closes the game (no further guesses allowed)
  closeGame () from Owner;
}
\end{lstlisting}

The \lstinline|StraightLineGuessingGame| protocol introduces, in the first line,
the roles users are expected to take when interacting with the smart contract:
the \lstinline|Owner| owns the game; the \lstinline|Player| makes guesses.
Stateful protocols require state to be kept in the contract. The
\lstinline|field|
declaration introduces the types of the fields that are stored within the state machine. In
this case, we need an \lstinline[morekeywords={HashedString}]|HashedString| for the secret. 
Along with the declared fields, \ourlanguage creates an extra
\lstinline[morekeywords={Value}]|Value| field by default, this field is used 
to manage the funds in the contract, 
in this instance, we use it to store the prize. 
The fields are stored in the state machine in the form of tuples, with each element of the
tuple corresponding to one of the declared fields. Users may declare repeated types. The tuple with 
stored contents can be used by the programmer when implementing the business logic. 
Protocol \lstinline|StraightLineGuessingGame| makes use of interaction constructs to describe interactions between an user and the endpoints \lstinline|lock|, \lstinline|guess| and \lstinline|closeGame|. In this protocol, the three endpoints
must be exercised once, in the order by which they appear in the protocol, and by users
of the appropriate role. Endpoint signatures comprise
the endpoint name followed by the types of the parameters.

It should be stressed that the guessing nature of the contract is nowhere
present in the protocol. Nothing in the protocol associates the
\lstinline|HashedString| in endpoint \lstinline|lock| to a secret, or
\lstinline|Value| to the prize. Nowhere it is said that guessing the secret
entails the transfer of the prize to the \lstinline|Player|'s account. Instead,
the protocol governs interaction only: which endpoints are available to which
roles, at which time. The business logic associated with the contract is
programmed later, in the contract language of the blockchain.

Our next version allows the owner to cancel the game after locking the secret
(perhaps the secret was too easy or the prize was set too low). The 
\lstinline[morekeywords={choice}]|choice| operator
denotes a choice made by an user, featuring different alternative
branches.

\begin{lstlisting}
protocol ChoiceGuessingGame (role Owner, role Player) {
  field HashedString; 
  lock (String, Value) from Owner;
  choice at Owner {
    proceedWithGame: { // owner wants players to guess
      guess (String) from Player;
    }
    cancelGame: { // the owner chooses to cancel the game    
    }
  }
  closeGame () from Owner;
}
\end{lstlisting}

After locking the secret, the \lstinline|Owner| is given two choices: to cancel the game
(\lstinline|cancelGame|) or to allow a player to make a guess
(\lstinline|proceedWithGame|). The two branches represent two different
endpoints in the contract. The \lstinline|choice| is in the hands of a single
role, \lstinline|Owner| in this case. This role should
exercise one endpoint or the other (but not both). Protocols for the two
branches are distinct. In the case of \lstinline|proceedWithGame|, endpoint
\lstinline|guess| is to be called by \lstinline|Player|. The \lstinline|cancelGame| branch 
is empty. In either case, the
\lstinline|Owner| is supposed to close the game after making the choice.
 
The third version allows one or more players to continue guessing while the game
is kept open by the owner. We make use of 
\lstinline|rec|-loops for the effect.

\begin{lstlisting}
protocol RecGuessingGame (role Owner, role Player) {
  field HashedString; 
  lock (String, Value) from Owner;
  rec Loop {
    choice at Owner {
    proceedWithGame : { // owner wants players to guess
      guess (String) from Player;
      Loop;
    }
    cancelGame : { // the owner wants to cancel the game    
    }
  }
  closeGame () from Owner;
}
\end{lstlisting}

The \lstinline|rec| constructor introduces a labelled recursion point.
In this case the protocol may continue at the recursion point by means of the
\lstinline|Loop| label. In any iteration of the loop the owner is called to
decide whether the game continues or not (perhaps the secret was found or
the owner got tired of playing). If she decides \lstinline|proceedWithGame|, a
player is given the chance of guessing (by calling endpoint
\lstinline|guess|) and the owner is called again to decide the faith of the
game.

Version three requires a lot of \lstinline|Owner| intervention: the continuation
of the game depends on her choice---\lstinline|proceedWithGame| or
\lstinline|cancelGame|---after each guess.
We want protocols able to terminate automatically, based on guess validation or
on the passage of time. Our fourth version takes advantage of the
\lstinline|do-interrupt| constructor and the \lstinline|trigger| declaration:

\begin{lstlisting}
protocol GuessingGame (role Owner, role Player) {
  field HashedString; 
  lock (String, Value) from Owner {
    // triggers for funds and slot
    funds trigger closeGame;
    slot trigger closeGame;
  };
  do {
    rec Loop {
      guess (String) from Player;
      Loop;
    }
  }
  interrupt {
    // close the game when one of the triggers is activated
    closeGame () from Contract;
  }
}
\end{lstlisting}

In the last version of the protocol, the \lstinline|Owner| does not have
any further involvement
after locking the secret. The game ends when one of
the triggers is activated.
Declarations \lstinline|slot trigger closeGame|
and \lstinline|funds trigger closeGame|
contain the keywords \lstinline|slot| and \lstinline|funds|, that instruct the compiler to generate 
functions in the business logic module
where the programmer may define the conditions for these triggers.
The primitive role \lstinline|Contract|
signs the \lstinline|closeGame| operation. This is not an endpoint, but an interaction 
that is executed automatically.

Constructors \lstinline|rec|, \lstinline|choice|, \lstinline|do-interrupt| and protocol definitions share Scribble's syntax entirely. Interactions are simplified: we remove \lstinline|to <recipient>| present in Scribble syntax because in our setting the recipient is always the contract. The declarations \lstinline|field|, \lstinline|funds trigger|, \lstinline|slot trigger| are exclusive to \ourlanguage.

%% file: contract_gen.tex
\section{Smart contract generation from \ourlanguage}
\label{sec:generation}

This section details the smart contracts generated from \ourlanguage protocols
and explain how developers can add custom business logic to complete contract
code. To ensure the validation of participants' interactions with the contract
we construct a finite state machine from each \ourlanguage protocol, whose
implementation is automatically generated by our compiler.

\begin{figure}[t!]
	\begin{subfigure}[b]{0.45\linewidth}
		\centering
		\includegraphics[width=0.75\linewidth]{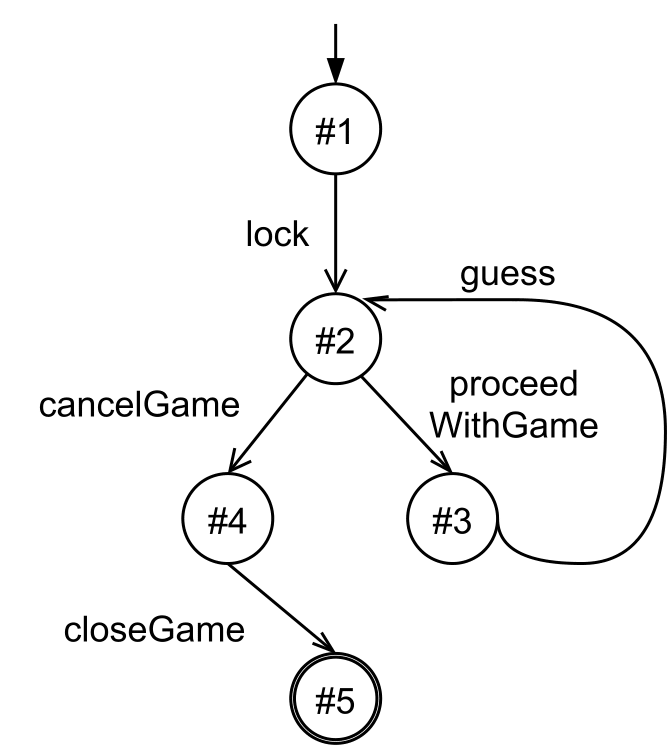}
		\caption{\lstinline|RecGuessingGame|}
		\label{fig:Automaton_ForRec}
	\end{subfigure}
	\hfill
	\begin{subfigure}[b]{0.45\linewidth}
		\centering
		\includegraphics[width=0.6\linewidth]{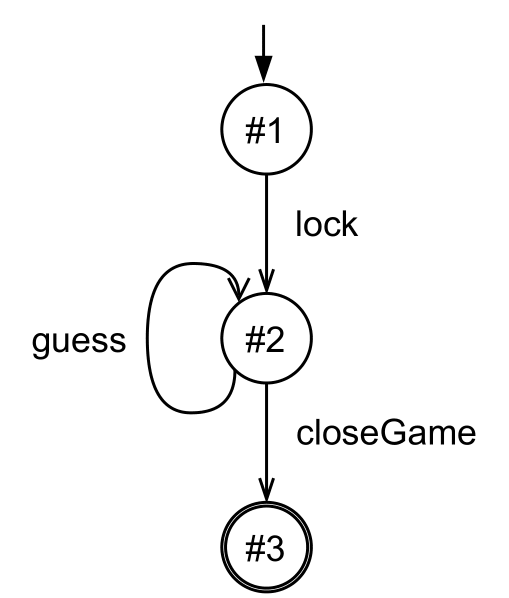}
		\caption{\lstinline|GuessingGame|}
		\label{fig:DoInterruptAutomaton}
	\end{subfigure}
	\caption{Finite state automata for two \ourlanguage protocols}
	\label{fig:automata}
	\vspace*{-2mm}
\end{figure}

Protocols are governed by finite state machines. \Cref{fig:automata} depicts the
automata for the recursive and the
\lstinline[morekeywords={do,interrupt}]|do-interrupt| guessing game in
\cref{sec:protocols}. The correspondence is such that endpoints in the automaton
correspond to edges, and pre- and post-interaction points in a protocol
correspond to nodes. Sequential composition, \lstinline|choice| and
\lstinline|rec| generate appropriate wiring. Interrupts call the associated
function generating new edges, as in the case of \lstinline|closeGame|
where an edge links state \#2 to the terminal state \#3
(\Cref{fig:DoInterruptAutomaton}).

The \ourlanguage compiler generates code divided in three different modules: \emph{Domain and Library Module}, \emph{Smart Contract Module} and \emph{Business Logic Module}. 
In this section we give a brief overview of the
three modules.

The \emph{Domain and Library Module} includes declaration of errors, the interface for the contract, the definition of state machine inputs, states and
the functions to interact with the state machine.

The \emph{Smart Contract Module} contains code activated when interacting with
endpoints. For example, \lstinline|lock| registers the two triggers and calls
the corresponding function in the Business Logic Module. The latter function
returns either an error or the fields to be stored at the new state. If an error
is received, no state transition is performed. Otherwise, the machine advances to
the next state and sets its new contents. Changes to the value field stored in
the state results in the transfer of funds between the node interacting with the endpoint
and the contract. 
This module also contains code for state transition that is used to define 
the state machine and boilerplate code
specific to Plutus' contracts.
To implement the state machine we use the Plutus State Machine library\footnote{\href{https://playground.plutus.iohkdev.io/tutorial/haddock/plutus-contract/html/Language-Plutus-Contract-StateMachine.html}{Language.Plutus.Contract.StateMachine}}, part of
the standard Plutus package.

Finally, the \emph{Business Logic Module} contains signatures for each of the
endpoints in the contract. The actual code is meant to be written by the
contract developer. The interaction
\begin{lstlisting}
lock (String, Value) from Owner;
\end{lstlisting}
in the protocol requires a function
\begin{lstlisting}[language=Haskell]
lock :: String -> Value -> Contract (Either StateContents Error)
\end{lstlisting}
(signature simplified) in the Smart Contract Module that returns either a new
\lstinline|StateContents| (a tuple composed of the fields stored in the state) 
or an error. If the value is non-positive, \lstinline|lock|
returns an error including an error message; otherwise returns a \lstinline|StateContents|,
that is a pair, composed of the hashed string corresponding to the input and the
value. These are the two fields to store in the new state of the state machine
(state\#2 in \Cref{fig:DoInterruptAutomaton}, reached via the
\lstinline|guess|-labelled edge).
The triggers: \lstinline|funds trigger| and \lstinline|slot trigger|
associated with \lstinline|lock|, generate
functions with signatures: 
\begin{lstlisting}[language=Haskell]
lockFundTrigger :: String -> Value -> Contract (Value -> Bool)
lockSlotTrigger :: String -> Value -> Contract Slot
\end{lstlisting}
for the respective triggers. In \lstinline|lockFundTrigger|, 
the developer should add an expression
with the condition to activate the trigger, e.g., 
\lstinline|(\funds -> funds `V.leq` 0)|. In \lstinline|lockSlotTrigger|
we specify the \lstinline|Slot| that activates the trigger.

Corresponding to the
\begin{lstlisting}[morekeywords={role,protocol,choice,rec,trigger,funds,slot,from,field}]
guess (String) from Player;
\end{lstlisting}
in the protocol, a function with the following signature must be written.
\begin{lstlisting}[language=Haskell]
guess :: String -> Contract (Either StateContents Error)
\end{lstlisting}

Function \lstinline|guess| reads the secret from the machine state (an
\lstinline|HashedString|) compares with the hashed version of the input string.
If they match, it returns a pair whose second component is zero ADA,
otherwise it returns an appropriate error message.
The caller to \lstinline|guess| (in module Smart Contract Module) detects the
difference in the value field of the state and credits the difference in the
client's account.
Finally, the
\lstinline|closeGame () from Contract| interaction point needs a function with the same name that, in
this case, returns a state with \lstinline|HashedString| \lstinline|"Game over"| and zero ADA as its fields.

The complete code of the module is in \cref{sec:plutus_code}; the code for the
three functions and two triggers amounts to 13 lines.

\begin{figure}[t!]
  \includegraphics*[width=0.9\linewidth]{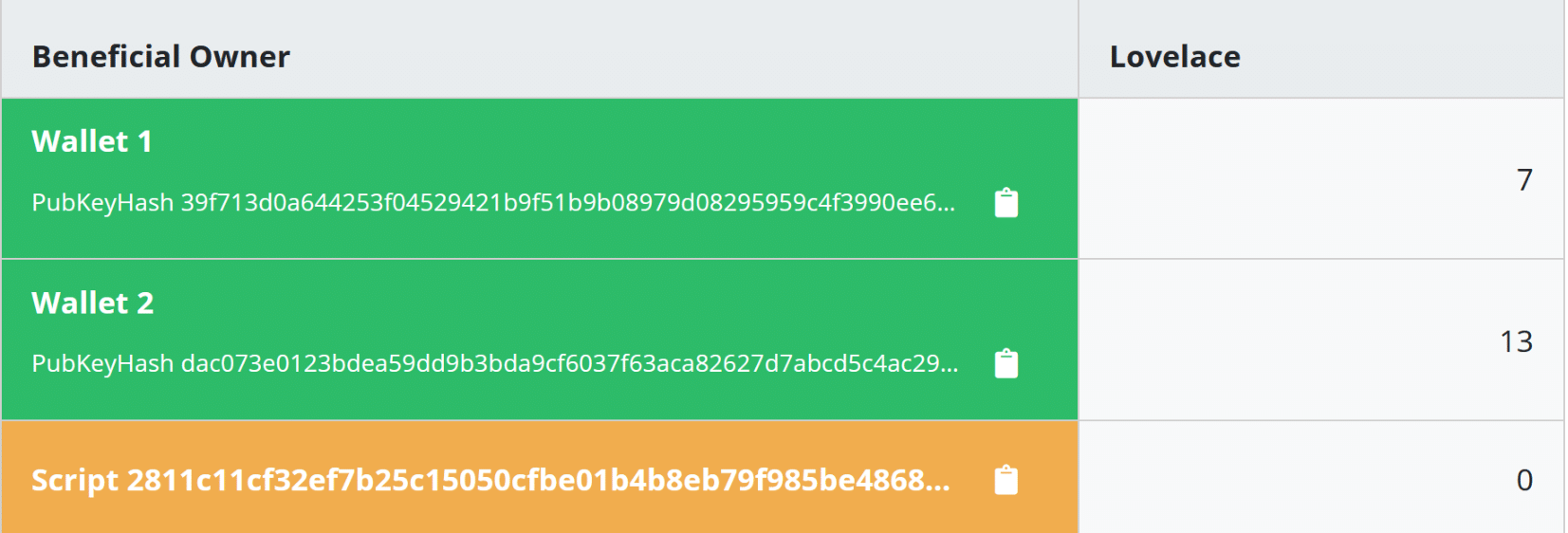}
  \caption{Final balances for setup in Fig.~\ref{Input_TwoLockPlayground_WithoutSM}, 
  using \ourlanguage (Plutus Playground)}
  \label{Output_TwoLockPlayground_WithSM_Balances}
  \vspace*{-3mm}
\end{figure}

We now revisit the scenario in \Cref{Two_Lock_GuessGame} using the input from \Cref{Input_TwoLockPlayground_WithoutSM}, that previously resulted in an error.
Thanks to the integration with a state machine, the contract now impedes the
second lock from taking effect, as can be seen in
\Cref{Output_TwoLockPlayground_WithSM_Logs} (available in~\cref{sec:logs}).
In \Cref{Output_TwoLockPlayground_WithSM_Balances}, we see that now the player is able to retrieve the prize for guessing correctly:
the player terminates with 13 Lovelace, and the owner ends with 7 Lovelace, resulting from the
deposit of 3 Lovelace with the first lock.

%% file: evaluation.tex
\section{Evaluation}
\label{sec:evaluation}

This section compares the amount of lines of code (LOC), those written by the developer
and those generated by our compiler. To carry out the comparison, we use the
guessing game together with three protocols representative of simple smart
contracts. The new protocols are in \cref{sec:ap_protocols}.

\begin{description}
\item[Ping Pong] A simple protocol that alternates between \lstinline|ping| and
  \lstinline|pong| operations, \textit{ad eternum}. No business logic is required for
  this protocol.
\item[Crowdfunding] A crowdfunding where an owner starts a campaign with a goal (in ADA), and 
	contributors donate to the campaign. When the owner decides to close the campaign,
	all the donations stored in the contract are collected.
\item[Auction] A protocol where a seller
  starts an auction over some token, setting the time limit and the maximum number of bids.
  Buyers bid for the token. When the auction is over, the seller collects
  the funds of the highest bid and the corresponding bidder gets the token.
\end{description}
 
\begin{table}[t!] 
  \centering
    \caption{Lines of code for each use case. From left to
  	right: LOC of the \ourlanguage protocol, LOC of the generated code, LOC of
  	the business logic, LOC 
  	for suggested implementation, ratio between LOC of the protocol and the
  	generated code, ratio between LOC written by the programmer (protocol +
  	business logic) and smart contract, ratio between LOC written by the developer
  	and suggested implementation}
  \begin{tabular}{| l | r | r | r | r | r | r | r |} 
    \hline
    \makecell{Protocol} & \makecell{Protocol \\ (LOC)} & \makecell{Generated \\ (LOC)} & \makecell{Logic \\ (LOC)} & \makecell{Suggested \\ (LOC)} & $\frac{\text{Protocol}}{\text{Generated}}$ & $\frac{\text{Written}}{\text{Contract}}$ & $\frac{\text{Written}}{\text{Suggested}}$ \\ [1.25ex] 
    \hline
    Ping Pong & 8 & 177 & 0 & 142 & 4.52\%  &  4.52\% & 5.63\% \\  
    \hline
    Crowdfunding & 12 & 187 & 6 & 163 & 6.42\% & 9.63\% & 11.04\% \\
    \hline
    Guessing Game & 16 & 162 & 13 & 165 & 9.88\%  & 17.90\% & 17.57\% \\
    \hline
    Auction & 15 & 150 & 22 & 185 & 10.00\% & 25.67\% &  20.00\%\\
    \hline 
  \end{tabular}
  \label{table:LOCComp}
\end{table}

\Cref{table:LOCComp} summarizes the analysis. Depending on the protocol, the
amount of generated code varies from 150 to 187 lines. In all our examples, the
generated code is at least $10 \times$ larger than the source written in
\ourlanguage. The business logic varies a lot from contract to contract;
nevertheless, it is important to note that it is extremely likely to be a small
portion of the complete contract due to the amount of necessary boilerplate that
Plutus requires. We see that the ratio between all the code written by the programmer 
(that is, the protocol and the business logic code) and the Plutus code that 
would otherwise be manually written is less than 1/4 in all analysed contracts. 
When we compare LOC for suggested implementations\footnote{Implementations available on \href{https://github.com/input-output-hk/plutus/tree/master/plutus-use-cases/src/Plutus/Contracts}{Plutus' GitHub}}
and \ourlanguage, we conclude that with \ourlanguage, the code manually written
is once again 1/5 or less for every scenario. Even in implementations developed
by experts, the boilerplate portion of the contract is significant.


%% file: conclusion_futurework.tex
\section{Conclusion and future work}
\label{sec:conclusion}

We present \ourlanguage---a protocol language for smart contracts---and a
compiler that automatically generates all contract code, except the business
logic. The generated code relies on state machines to prevent
unexpected interactions with the contract.
We claim that \ourlanguage improves the reliability of contracts by reducing the
likelihood of programmers introducing faults in smart contract code. Our 
language also flattens the learning curve,
allowing developers to focus on the business logic rather than on the
boilerplate required to setup a contract, namely in Plutus.
Preliminary results  point to a 1/4 ratio between the number of lines of code
written by the programmer and those in the final contract.
This paper constitutes an initial report on using protocol descriptions to generate
contract code. Much remains to be done. \ourlanguage protocols classify
participants under different roles, but we currently do not enforce any form of
association of participants to roles. We plan to look into different forms of
enforcing the association.
Business logic is currently manually written in the contract language (Plutus)
and added to the code generated from the protocol. We plan to look into ways of
adding more business logic to protocols, thus minimising the Plutus code that
must be hand written.
Some features of \ourlanguage are strongly linked with Plutus. The
trigger generation is one of those features: it depends on Plutus libraries for the effect. Nevertheless, we 
believe that \ourlanguage can be adapted to target other languages with minimal changes
to the syntax and semantics. 
Generating Solidity code might be an interesting option for the future, as it also supports
state machines.
Lastly, evaluation needs to be elaborated. In ongoing work, 
we are comparing the usage of computational 
resources between contracts implemented with \ourlanguage and the 
corresponding suggested implementations.

%% file: guessing-game-code.tex
\section{Plutus code for vulnerable guessing game}
\label{sec:guessgame_code}
\begin{lstlisting}[language=Haskell,tab=2,showstringspaces=false,breaklines=true]
module GuessingGame 
  ( guess
  , lock
  , endpoints)
  where

import           Control.Monad             (void)
import qualified Data.ByteString.Char8     as C
import           Language.Plutus.Contract
import qualified Language.PlutusTx         as PlutusTx
import           Language.PlutusTx.Prelude hiding (pure, (<$>))
import           Ledger                    (Address, Validator, ValidatorCtx, Value, scriptAddress)
import qualified Ledger.Constraints        as Constraints
import qualified Ledger.Typed.Scripts      as Scripts
import           Playground.Contract
import qualified Prelude

newtype HashedString = HashedString ByteString deriving newtype PlutusTx.IsData

PlutusTx.makeLift ''HashedString

newtype ClearString = ClearString ByteString deriving newtype PlutusTx.IsData

PlutusTx.makeLift ''ClearString

type GameSchema =
    BlockchainActions
        .\/ Endpoint "lock" LockParams
        .\/ Endpoint "guess" GuessParams

data Game
instance Scripts.ScriptType Game where
    type instance RedeemerType Game = ClearString
    type instance DatumType Game = HashedString

gameInstance :: Scripts.ScriptInstance Game
gameInstance = Scripts.validator @Game
    $$(PlutusTx.compile [|| validateGuess ||])
    $$(PlutusTx.compile [|| wrap ||]) where
        wrap = Scripts.wrapValidator @HashedString @ClearString

-- create a data script for the guessing game by hashing the string
-- and lifting the hash to its on-chain representation
hashString :: String -> HashedString
hashString = HashedString . sha2_256 . C.pack

-- create a redeemer script for the guessing game by lifting the
-- string to its on-chain representation
clearString :: String -> ClearString
clearString = ClearString . C.pack

-- the validation function (Datum -> Redeemer -> ValidatorCtx -> Bool)
validateGuess :: HashedString -> ClearString -> ValidatorCtx -> Bool
validateGuess (HashedString actual) (ClearString guess') _ = actual == sha2_256 guess'

-- the validator script of the game.
gameValidator :: Validator
gameValidator = Scripts.validatorScript gameInstance

-- the address of the game (the hash of its validator script)
gameAddress :: Address
gameAddress = Ledger.scriptAddress gameValidator

-- parameters for the "lock" endpoint
data LockParams = LockParams
    { secretWord :: String
    , amount     :: Value
    }
    deriving stock (Prelude.Eq, Prelude.Show, Generic)
    deriving anyclass (FromJSON, ToJSON, IotsType, ToSchema, ToArgument)

--  parameters for the "guess" endpoint
newtype GuessParams = GuessParams
    { guessWord :: String
    }
    deriving stock (Prelude.Eq, Prelude.Show, Generic)
    deriving anyclass (FromJSON, ToJSON, IotsType, ToSchema, ToArgument)

-- the "lock" endpoint. 
lock :: AsContractError e => Contract GameSchema e ()
lock = do
    LockParams secret amt <- endpoint @"lock" @LockParams
    let tx         = Constraints.mustPayToTheScript (hashString secret) amt
    void (submitTxConstraints gameInstance tx)

-- the "guess" endpoint. 
guess :: AsContractError e => Contract GameSchema e ()
guess = do
    GuessParams theGuess <- endpoint @"guess" @GuessParams
    unspentOutputs <- utxoAt gameAddress
    let redeemer = clearString theGuess
        tx       = collectFromScript unspentOutputs redeemer
    void (submitTxConstraintsSpending gameInstance unspentOutputs tx)

game :: AsContractError e => Contract GameSchema e ()
game = lock `select` guess

endpoints :: AsContractError e => Contract GameSchema e ()
endpoints = game

mkSchemaDefinitions ''GameSchema

$(mkKnownCurrencies [])
\end{lstlisting}

%% file: logs.tex
\section{Plutus playground simulation for the guessing game scenario} \label{sec:logs}

\begin{figure}[H]
	\centering
	\includegraphics*[width=0.8\linewidth]{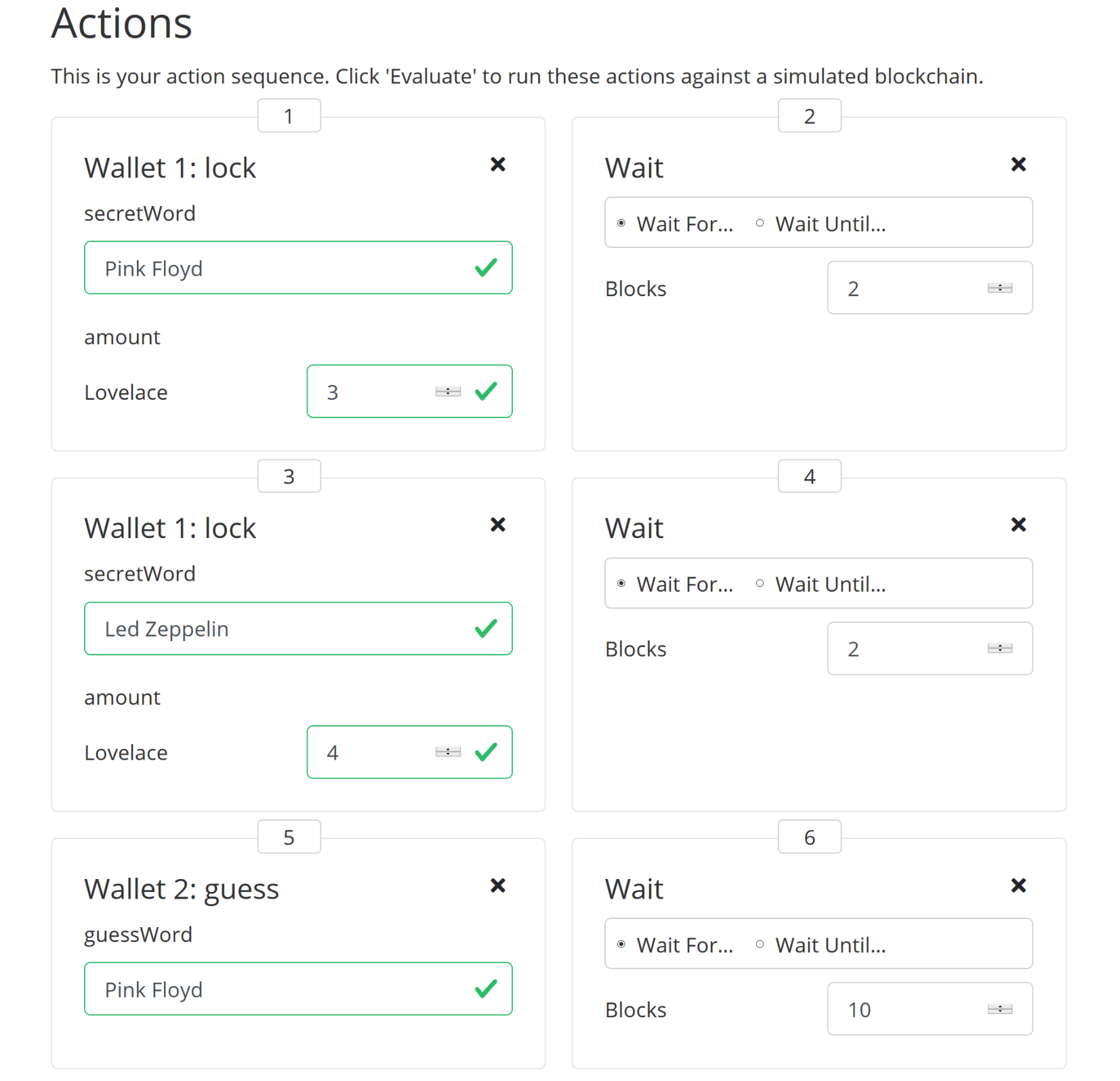}
	\caption{Complete version of Fig.~\ref{Input_TwoLockPlayground_WithoutSM}. This is the setup to run the scenario illustrated in Fig.~\ref{Two_Lock_GuessGame} in Plutus Playground: the owner (Wallet 1) makes two consecutive locks (``Pink Floyd'' and ``Led Zeppelin''), the player (Wallet 2) guesses the first secret. Wait in between actions, to let the simulator process the inputs.}
	\label{Input_TwoLockPlayground_Complete}
\end{figure}

\begin{figure}[H]
	\texttt{
		=== Slot 1 ===\\
		Wallet 1: \\
		ReceiveEndpointCall ("tag":"lock","value":3,"secretWord":"Pink Floyd")\\
		Validating transaction: [..]\\
		=== Slot 2 ===\\
		=== Slot 3 ===\\
		Wallet 1: \\
		ReceiveEndpointCall ("tag":"lock","value":4,"secretWord":"Led Zeppelin")\\
		Validating transaction: [..]\\
		=== Slot 4 ===\\
		=== Slot 5 ===\\
		Wallet 2:\\
		ReceiveEndpointCall ("tag":"guess","guessWord":"Pink Floyd")\\
		{\color{red}   Validation failed: [..] (ScriptFailure (EvaluationError []))}
	}
	\caption{Log resulting from the setup in
		Fig.\ref{Input_TwoLockPlayground_WithoutSM} with a \textit{vanilla} Plutus smart contract; observe that both locks are
		validated (\texttt{Slot 1} and \texttt{Slot 3}) and that a correct guess later
		results in a validation failure (\texttt{Slot 5})}
	\label{Output_TwoLockPlayground_WithoutSM_Logs}
\end{figure}

\begin{figure}[H]
	\texttt{Validating transaction: [...]\\
		=== Slot 1 ===\\
		Wallet 1:\\
		EndpointCall ("tag":"lock","value":3,"secretWord":"Pink Floyd")\\
		Validating transaction: [...]\\
		=== Slot 2 ===\\
		Wallet 1: "No previous state found, initialising SM with state: \\
		LockState (HashedString ...) and with value: 3"\\
		=== Slot 3 ===\\
		Wallet 1:\\
		EndpointCall ("tag":"lock","value":4,"secretWord":"Led Zeppelin")\\
		Wallet 1: {\color{red} "Previous lock detected. \\
			This lock produces no effect"}\\
		=== Slot 4 ===\\
		=== Slot 5 ===\\
		Wallet 2: \\
		EndpointCall ("tag":"guess","guessWord":"Pink Floyd")\\
		Wallet 2: "Congratulations, you won!"\\
		Validating transaction: [...]\\
		=== Slot 6 ===\\
		Wallet 1: "Closing the game"\\
		Wallet 2: \\
		"Successful transaction to state: LockState (HashedString ...)"\\
		Validating transaction: [...]\\
		=== Slot 7 ===\\
		Wallet 1:  \\
		"Successful transaction to state: CancelGameState (HashedString ...)"
	}
	\caption{Log resulting from the setup in
		Fig.\ref{Input_TwoLockPlayground_WithoutSM}. This time using the contract
		generated from the protocol using \ourlanguage; observe that the second lock fails (\texttt{Slot
			3}) and therefore the guess from Wallet 2 is successfully validated}
	\label{Output_TwoLockPlayground_WithSM_Logs}
\end{figure}

%% file: scribble-protocols.tex
\section{\ourlanguage protocols for the evaluation section}
\label{sec:ap_protocols}

\noindent\textbf{The ping-pong protocol}

\begin{lstlisting}
protocol PingPongRec (role Client) {
  init() from Client;    
  rec Loop {
    ping() from Client;    
    pong() from Client;    
    Loop;
  }    
}
\end{lstlisting}

\noindent\textbf{The crowdfunding protocol}

\begin{lstlisting}
protocol Crowdfunding (role Contributor, role Owner){
	init (Value) from Owner;
	rec Loop {
		choice at Owner{
			continue : {
				contribute (Value) from Contributor; 
				Loop;
			}
			closeCrowdfund : {}
		}
	}
}
\end{lstlisting}

\noindent\textbf{The auction protocol}

\begin{lstlisting}
protocol Auction (role Seller, role Buyer) {
  field PubKeyHash, Value;
  beginAuction (Token, Value) from Seller {
    slot trigger (slot == 10, endAuction);
  };
  do {
    rec Loop {
      bid (Value) from Buyer;
      Loop;
    }
  }
  interrupt {
    endAuction () from Contract;
  }
}
\end{lstlisting}

%% file: plutus-code.tex
\section{Plutus code for business logic}
\label{sec:plutus_code}

\noindent\textbf{Guessing game logic module (\ourlanguage)}
\begin{lstlisting}[language=Haskell,tab=2,showstringspaces=false,breaklines=true]
module GuessingGameLogic
  ( lock
  , guess
  , closeGame
  ) where

import           Control.Lens
import           Control.Monad            (void)
import           GHC.Generics             (Generic)
import           Language.PlutusTx.Prelude
import qualified Ledger.Ada               as Ada
import qualified Ledger.Value             as V
import qualified Data.Text                as T
import qualified Data.ByteString.Char8    as C
import qualified GuessGameLibrary         as G
import PlutusTx.Prelude
import qualified Prelude

hashString :: String -> HashedString
hashString = HashedString . sha2_256 . C.pack

zeroLovelace :: Value
zeroLovelace = Ada.lovelaceValueOf 0

lockFundTrigger :: G.AsError e => String -> Value -> Contract G.Schema e (Value -> Bool)
lockFundTrigger str val = 
    pure $ (\presentVal -> presentVal `V.leq` zeroLovelace)

lockSlotTrigger :: G.AsError e e => String -> Value -> Contract G.Schema e Slot
lockSlotTrigger param1 param2 = pure $ 10

lock :: G.AsError e => String -> Value -> Contract G.Schema e (Either G.State G.Error)
lock str val = pure $
  if  val `V.leq` zeroLovelace 
    then Right $ Error $ T.pack $ "The prize must be greater than 0; got " <> show val
    else Left (hashString str, val)

guess :: G.AsError e => String -> Contract G.Schema e (Either G.State G.Error)
guess str = do
    secret <- mapError (review _GuessingGameSMError) $ G.getCurrentStateSM G.client
    if secret == hashString str
      then do
        logInfo @String "Congratulations, you won!"
        pure $ Left (hashString str, zeroLovelace)
      else pure $ Right $ Error "Incorrect guess, try again"

closeGame :: G.AsError e => Contract G.Schema e (Either G.State G.Error)
closeGame = do 
  logInfo @String "Closing the game"
  pure $ Left (hashString "Game over", zeroLovelace)

\end{lstlisting}
